\documentclass{aastex}
\input{epsf.sty}

\shorttitle{Effects of DM profiles on bulge formation}
\shortauthors{Huang, Deng \& Fu}

\begin{document}


\title{Effects of dark-matter density profiles on bulge
formation in very late-type galaxies}


\author{Jie-Hao Huang}
\affil{Department of Astronomy, Nanjing University, Nanjing
210093, China} \email{jhh@nju.edu.cn}

\author{Zu-Gan Deng}
\affil{Department of Physics, Graduate School, Chinese Academy of
Sciences, Beijing 100039, China} \email{dzg@vega.pku.bac.ac.cn}

\and

\author{Yan-Ning Fu}
\affil{Purple Mountain Observatory, Chinese Academy of Sciences,
Nanjing 210008, China}

\affil{National Observatories, Chinese Academy of Sciences,
Beijing 100039, China} \email{yanningfu@yahoo.com}




\begin{abstract}
The dynamical evolution of super star clusters has been
investigated in dark matter halos depicted with a cuspy- or
soft-core density profile. The simulations show that (1)
exponential bulges with central cusps form in both cases;(2)
distinctive bulge formation rates are derived for cuspy- and
soft-core profile; (3) masses of bulges and nuclear clusters are
more heavier in case of cuspy-core profile; (4) massive, nuclear
star clusters possibly form with no discernible bulges at
different ages in cuspy- or soft-core cases.
\end{abstract}


\keywords{galaxies: bulge -- galaxies: kinematics and dynamics --
star: super star clusters}


\section{Introduction}

In our previous work (Fu, Huang \& Deng 2003a, (FHD03)), we
managed to construct a set of models for simulating the dynamical
evolution of super star clusters (SSCs) embedded in dark matter
(DM) halo with central cusp, motivated by observations in nuclear
regions of galaxies. Among these observations we would mention the
discovery of the presence of a black hole with intermediate mass
(IMBH) in a young SSC located at some distance from the nucleus of
M82 (Matsumoto et al. 2000). While another possible IMBH has been
found in an evolved SSC located in the nucleus of M33 (Gebhardt et
al 2001). These discoveries strongly imply the role of dynamical
evolution of SSCs in the physical process causing the $M_{\rm BH}
-  M_{\rm bulge}$ relation (see, Fu, Huang \& Deng 2003b), as well
as in bulge formation in late-type galaxies.

The simulations in FHD03 have yielded bulges that are similar in
many aspects to the observational ones. In particular, the derived
surface density profiles can be well fitted by an exponential
structure with nuclear cusps, which is consistent with {\it Hubble
Space Telescope (HST)} observations (Carollo 1999). The progress
in observations (e.g. Marchesini et al. 2002; and the references
therein) shows, however, that the cuspy-core DM profiles, such as
the NFW profile, are not adequate for a large fraction of dwarf
galaxies which are dominated by the DM halos, and suggested that
the soft-core density profiles, the Burkert profile for example,
is more suitable to these galaxies. It would then be very much
instructive to perform simulations for the dynamical evolution of
SSCs with both the cuspy- and the soft-core DM density profiles
for the bulge formation, with the hope of constraining the
bulge-formation model or the properties of the dark matter in
late-type galaxies. That is what we would present in this paper.

\section{Models}

In view of what we described in introduction, we would then model
the sinking and tidal stripping of SSCs in DM halos depicted with
the NFW or the Burkert density profiles in this paper so as to
derive something important as compared with observations. The part
of the models in FHD03 are adjusted accordingly and are described
in the following subsections. The readers are referred to FHD03
for detailed description about the other part of the models which
remain unchanged in this work, including dynamical friction and
tidal stripping.

\subsection{SSC model}
Recently, Parodi \& Binggeli (2003) studied the spatial
distribution of bright star-forming complexes in a volume-limited
sample of 72 very late-type  galaxies located within  10 Mpc. They
derived  the projected radial number distribution $R
\exp(-R/R_l)$, with median scale length $R_l \approx 0.6kpc$ and
median number of bright star-forming lumps 17, respectively.
Taking this spatial distribution with its median values  as our
SSCs' radial distribution, rather than {\em 1/R} function inferred
from 3 galaxies in FHD03, we generate a set of SSCs over the whole
galaxy instead of setting both inner and outer boundary for its
distribution.   As in FHD03, these SSCs are assigned with local
circular speed. Both the initial mass function of SSCs and the
SSC itself remain unchanged as  those in FHD03.

\subsection{Background}

Since  very-late type galaxies are DM dominated at all radii
\citep{mar02,per96}, we only take, as in FHD03, a DM halo as the
initial background for the bulge-formation process. In light of
considering the effects of DM halo models on the bulge formation,
we adopt the NFW \citep{nav97} and the Burkert (1995) profiles as
the cuspy- and the soft-core density profiles, respectively.  The
mass of the dark matter halo is taken to be $M_{200}=10^{11}
M_{\odot}$.

The NFW profile  can be written as \citep{bin98}
\begin{equation}
\label{NFWMassdensity} \rho_{NFW}(r)=\frac{M_{0h}}{r(a_h+r)^2},
\end{equation}
where $r$ is the distance from the halo center, and, $a_h$ and
$M_{0h}$ are two parameters. With a $\Lambda$CDM cosmological
model \citep{jin00}, the NFW halo is scaled to $M_{200}=10^{11}
M_{\odot}$, which gives
\begin{equation}
\label{NFWpa} a_h=9.1kpc, \ \ M_{0h}= 5.3 \times 10^{10}
M_{\odot}.
\end{equation}

The Burkert profile writes
\begin{equation}
\label{BurkertMassdensity} \rho_{Burkert}(r)=\frac{\rho_0
r_0^3}{(r+r_0)(r^2+r_0^2)},
\end{equation}
where the central density $\rho_0$ is taken to be
$0.05M_{\odot}/pc^3$ following Marchesini et al. (2002), and the
scale radius $r_0$ of 4 kpc is computed from the definition of
$M_{200}$.

\section{Results and Discussions}

\subsection{Properties of formed bulges and formation rates}

Fig~\ref{SurfDvsProjD} shows the evolution of surface density
profiles of simulated individual galaxies obtained with the NFW
and the Burkert density profiles, respectively. Obviously, the
displayed profiles in Fig 1 fall into two categories, one with
high-density central component and the other without. The latter
ones basically correspond to the cases where no SSC contributes
its mass to the bulge area, or to the cases of being at the
pre-stage of bulge-forming.  On the contrary, the former ones
correspond to the cases where there is some mass contribution from
SSCs down to less than 1 pc from the halo center, and the profiles
clearly show that bulges are well-formed.

The mean surface density profiles of the formed bulges derived
with the NFW or the Burkert profiles at 3 Gyr are illustrated in
Fig~\ref{MeanDvsProjD}, together with the curves depicting the
fitting to the model (Carollo \& Stiavelli 1998; FHD03),

\begin{equation}
\label{model} \sigma(R)=\sigma_0 \exp(-1.678
\frac{R}{R_e})+\sigma_1(1+\frac{R_c}{R})^\gamma
\exp(-\frac{R}{R_s}).
\end{equation}

As obtained in FHD03, the formed bulges, either  with the NFW or
the Burkert density profiles, do share the same characteristics
with the {\it HST} observational results (Carollo 1999), i.e. the
general presence of central cusps on top of the exponential
bulges.

The half-mass radii, $R_e$, of the formed exponential bulges for
individual galaxies, derived from the fitting with equa
(\ref{model}), range over 100pc to 1000pc.
Fig~\ref{NvsRevsFWHM}a,b show the histograms of the scale radii
$R_e$ of individual bulges formed at 3 Gyr. Either with the NFW or
the Burkert density profile, the scale radii $R_e$ distributed
over a few hundred parsecs are in accord with observed results by
the {\it HST} (Carollo 1999).

The above analyses indicate that neither the NFW nor the Burkert
profile make a difference to the formed bulges in view of their
structure. Nevertheless,  the fraction of simulated galaxies with
no bulges formed turn out to be quite different as
Fig~\ref{SurfDvsProjD} illustrated. This situation is clearly
manifested in Fig~\ref{FormRate}. The bulge formation rates are
quite low in simulated galaxies derived with the Burkert density
profile, about 30\% and 70\% at the age of 1 and 4 Gyr,
respectively. On the contrary, the formation rates derived with
the NFW profile are above 97\% and about 100\% at 1 and 4 Gyr,
respectively.

Observations on searching for bulges in very late-type galaxies
are at the very beginning indeed (see, e.g. B\"{o}ker et al. 2003;
Carollo 1999). These researches provide the bulge detection rates
of about 50\%, compatible with the bulge formation rates derived
from the simulations with the Burkert profile. Though it would be
natural to conclude  that the DM density profiles suitable to very
late-type dwarfs are soft-core ones, we should be cautious at the
moment. For one thing, these surveys are based on samples of small
number of sources, 19 galaxies at the most. On the other hand,
Matthews \& Gallagher (1997) found that bulges of extremely
late-type spirals are often very weak.


\subsection{The mass relation  of bulges to their nuclear cusps}
Though Carollo et al. (2002) discovered that each exponential
bulge in their sample hosts a stellar nucleus in its center, they
claimed no necessarily causal connection between these two
components. In our simulations (FHD03, and this work), however,
every formed bulge is always paired with its central, stellar
cusp. The stripped stars from SSCs contribute to the bulge
formation, while the remaining, undissociated  SSCs form the
nuclear cusps. Needless to say, to confirm the causal connection
between nuclear cusps and the exponential bulges would shed much
light on the evolution of galaxies.

Recent work by Balcells et al. (2003) seems to confirm the
existence of the causal connection between the paired bulges and
nuclear cusps, where a tight luminosity correlation  in H-band has
been reported. In order to see if there is a corresponding
correlation between the masses of our simulated bulges and their
paired cusps, we have measured the scale sizes of nuclear cusps
quantified by their FWHMs, following Carollo et al. (2002).
Considering the minimum angular size of 0.\arcsec 03 that WFPC2 on
board {\it HST} can resolve and the distance ranges of 12 and 37
Mpc for observed galaxies, we have averaged the simulated nuclear
cusps over central radius of 2 or 5 pc so that the measured FWHMs
of simulated nuclear cusps are comparable to the observed ones.
Fig~\ref{NvsRevsFWHM}c,d show the FWHMs thus obtained of the
simulated nuclear cusps, which conforms with observed results
(Carollo et al. 2002), ranging from a few pc up to 20 pc.

The derived mass relation between simulated bulges and their
paired nuclear cusps is illustrated in Fig~\ref{MCusvsMBul}, where
the correlations are obviously reached either for the NFW or the
Burkert profile at 3 Gyr. The consistency between the observed and
simulated mass correlation provides further evidence favorable to
the proposed scenario (FHD03 and this work) for bulge formation in
very late-type galaxies.

Besides, it would be worth noticing what Fig~\ref{MCusvsMBul}
reveals that the formed bulges and nuclear cusps are more heavier
in case of the NFW density profile, with median values of about
$2\times 10^{7}M_{\odot}$ and $1.5\times 10^{6}M_{\odot}$
respectively at 3 Gyr. On the other hand, the relevant values
derived in the Burkert profile are about $8\times 10^{6}M_{\odot}$
and $6\times 10^{5}M_{\odot}$ respectively.
This difference is mainly made by the different local densities
and density variation of DM within the circumnuclear region in the
cuspy- and soft-core profiles. Further observations are surely
needed to confirm this kind of disparity.

\subsection{Formation of nuclear star clusters without bulges}
Now it is believed that the occurrence of compact star clusters at
nuclei of very late-type spirals is a common phenomenon (e.g.
B\"{o}ker et al. 2002). A fraction of these clusters are located
in bulge-less galaxies, with ages roughly ranging from $10^{7}$ to
$10^{9} yr$ (e.g. Walcher et al. 2003).

As we have shown (FHD03 and this work), our proposed scenario
provides a promising mechanism for the formation of massive,
nuclear star clusters in galaxies having detected bulges, with
masses compatible with observed values ranging very crudely from a
few times $10^{5}$ to $10^{7}M_{\odot}$ (e.g. Matthews \&
Gallagher 2002).

The existence of nuclear star cluster in bulgeless galaxies,
however, remains a mystery. This phenomenon is closely related to
several hot topics, especially the properties of DM halos in
galaxies. We have then made records of simulations for spatial
distances of sinking SSCs during 1Myr and 3 Gyr, either in case of
the NFW or Burkert profile, to see if some massive SSCs could
survive tidal stripping so as to sink into center of simulated
galaxies before bulge formed.

\begin{figure*}
{\epsfxsize= 12 cm \epsfbox{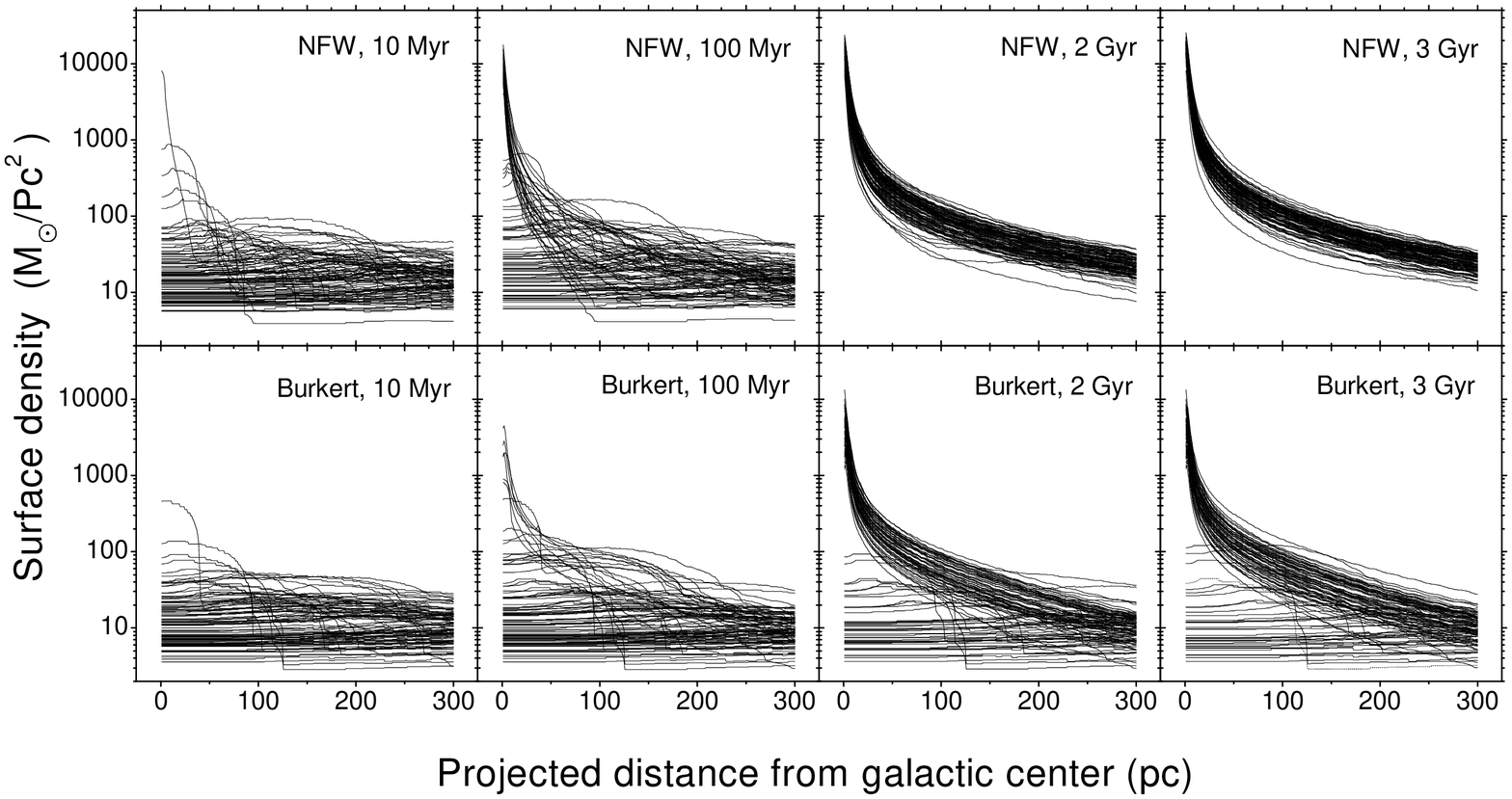}} \vspace{0cm}
\caption{Evolution of surface density profiles of simulated
galaxies in the NFW and Burkert profile. \label{SurfDvsProjD}}
\end{figure*}

\begin{figure*}
{\epsfxsize= 10 cm \epsfbox{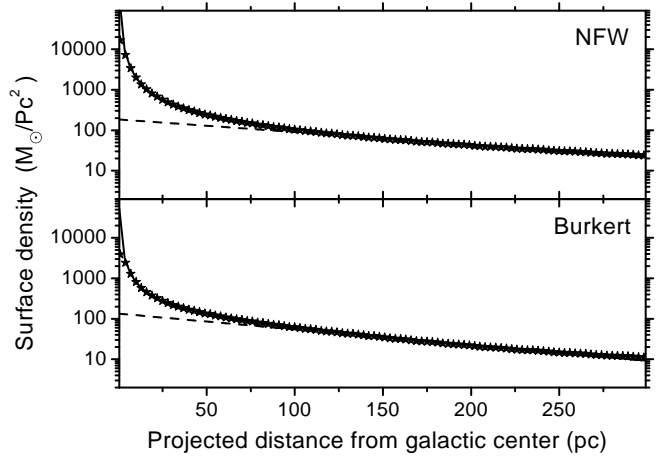}} \vspace{0cm} \caption{Mean
surface density profiles of the formed bulges, along with the
curves depicting the model fitting. Legends: Stars -- simulated
mean profiles, dashed lines -- fitted exponential components,
solid lines --  fitting to model (\ref{model}).
\label{MeanDvsProjD}}
\end{figure*}

\begin{figure*}
\vspace{-1cm}{\epsfxsize= 10 cm \epsfbox{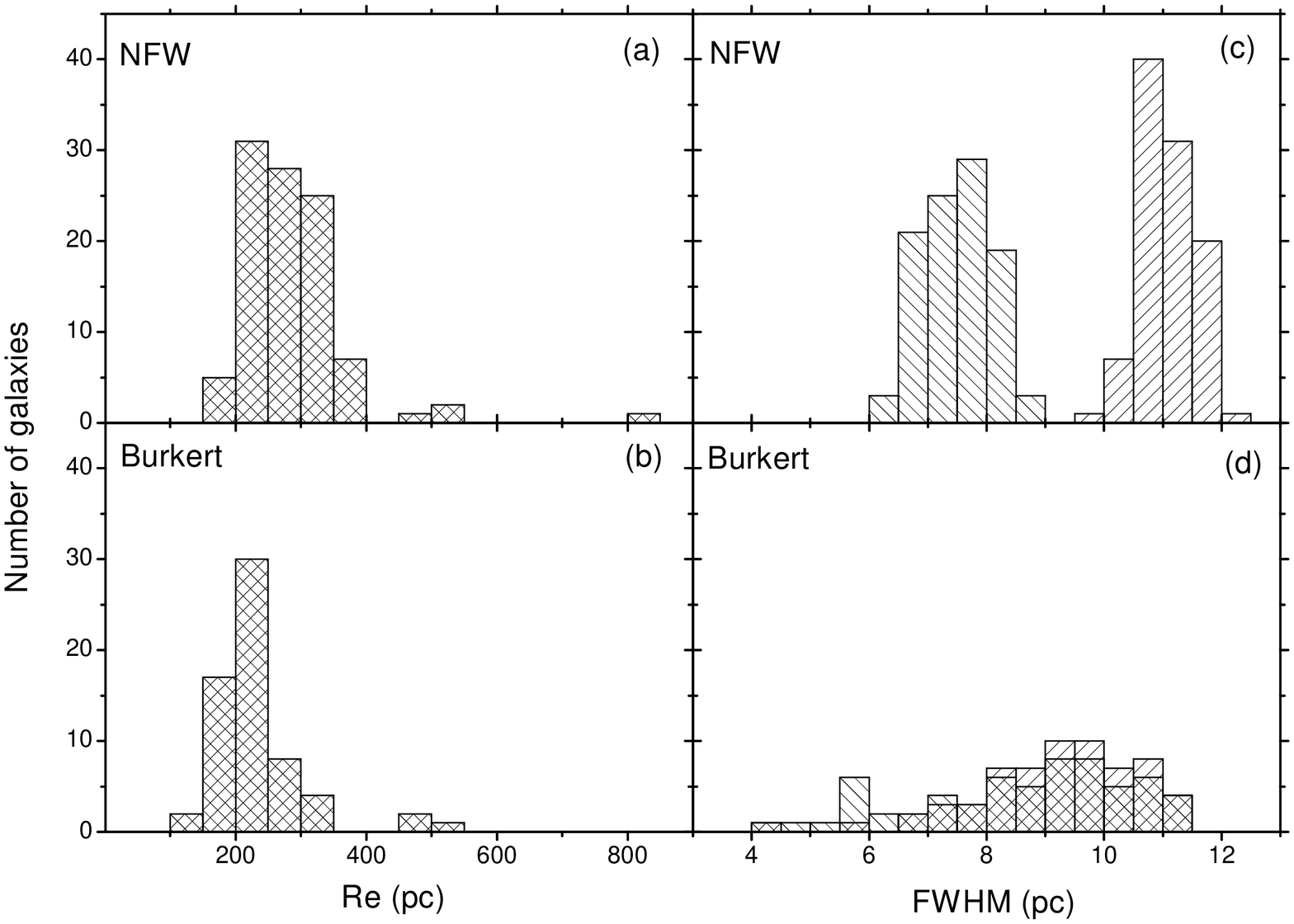}} \caption{(a) and
(b) Distribution of half-mass radii of individual bulges, $R_e$.
(c) and (d) FWHMs of simulated nuclear cusps. Legends: Columns
shaded with backward and forward slashes are results obtained by
setting the central peak values of surface density as those
averaged over central 2 and 5 pc globes, respectively. See the
text for details.\label{NvsRevsFWHM}}
\end{figure*}

\begin{figure*} \vspace{-2cm}{\epsfxsize= 10 cm
\epsfbox{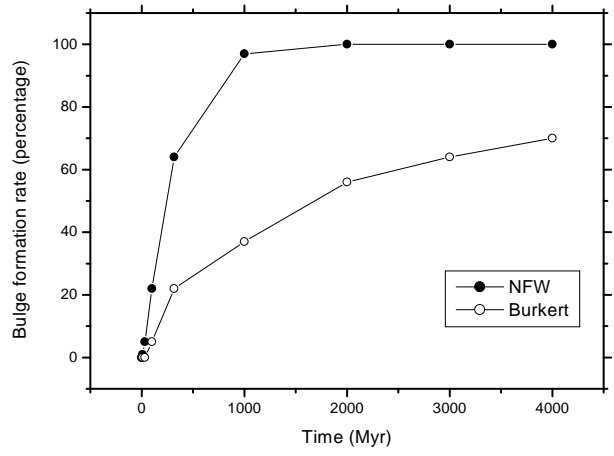}} \caption{Bulge formation rates in the NFW and
the Burkert profiles, respectively. \label{FormRate}}
\vspace{0cm}\end{figure*}

\begin{figure*}
{\epsfxsize= 10 cm \epsfbox{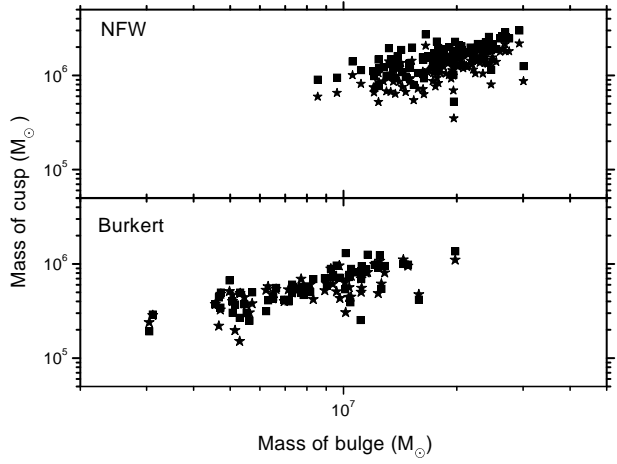}} \caption{Mass relation of the
formed bulges to their central cusps. Legends: The filled squares
and stars denote masses of central cusps obtained by spatial
resolution of 2 and 5 pc as Fig~\ref{FormRate}, respectively.
\label{MCusvsMBul}}

\end{figure*}

\begin{figure*}
{\epsfxsize= 12 cm \epsfbox{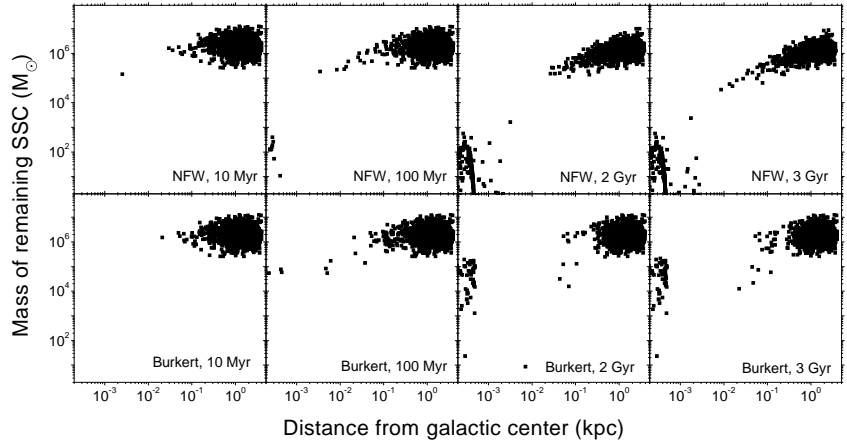}} \vspace{0cm} \caption{Mass of
remaining SSCs of simulated galaxies plotted against the SSCs'
distance from the halo center in the NFW and Burkert halos.
\label{MsscvsSpatD}}
\end{figure*}

The snapshots of this kind are illustrated in
Fig~\ref{MsscvsSpatD} for DM of NFW and Burkert profile,
respectively. It is very intriguing to find from
Fig~\ref{MsscvsSpatD} that massive, nuclear star clusters do form
in NFW profile at 10Myr and in Burkert profile around 100Myr with
no discernible bulges. Indeed, this situation has already been
manifested in Fig~\ref{SurfDvsProjD} at corresponding ages. After
10Myr, the probability to have nuclear star cluster with no bulge
formed becomes low for DM depicted with NFW density profile, while
it is low before around 10Myr in case of Burkert profile. After 1
Gyr, this probabilities are quite low in both cases, especially in
case of NFW profile due to very high bulge formation rate, above
97\%.

It seems then the scenario we proposed could be a promising
mechanism to form nuclear star clusters in bulge-less galaxies.
Before drawing a conclusion about the suitability of cuspy-core
density profiles for very late-type dwarfs, however, the models
presented in FHD03 do need inclusion of more physical processes,
e.g. the reaction of DM halo to the sinking of SSCs. This process
might have strong effect on changing the DM density profiles
(El-Zant et al. 2001).

\acknowledgements{This work is supported  by NKBRSF G19990754, and
NSFC.}

\clearpage

\clearpage

\clearpage

\clearpage

\clearpage

\clearpage

\end{document}